\documentclass{ws-procs975x65}
\usepackage{graphicx}

\begin{document}
\def\AJ{{\it Astron. J.} }
\def\ARAA{{\it Annual Rev. of Astron. \& Astrophys.} }
\def\ApJ{{\it Astrophys. J.} }
\def\ApJL{{\it Astrophys. J. Letters} }
\def\ApJS{{\it Astrophys. J. Suppl.} }
\def\ApP{{\it Astropart. Phys.} }
\def\AA{{\it Astron. \& Astroph.} }
\def\AAR{{\it Astron. \& Astroph. Rev.} }
\def\AAL{{\it Astron. \& Astroph. Letters} }
\def\JGR{{\it Journ. of Geophys. Res.}}
\def\JPhG{{\it Journ. of Physics} {\bf G} }
\def\PhFl{{\it Phys. of Fluids} }
\def\PR{{\it Phys. Rev.} }
\def\PRD{{\it Phys. Rev.} {\bf D} }
\def\PRL{{\it Phys. Rev. Letters} }
\def\PLB{{\it Phys. Lett. B}}
\def\Nature{{\it Nature} }
\def\NewAR{{\it New Astron. Rev.}}
\def\MNRAS{{\it Month. Not. Roy. Astr. Soc.} }
\def\ZA{{\it Zeitschr. f{\"u}r Astrophys.} }
\def\ZFN{{\it Zeitschr. f{\"u}r Naturforsch.} }
\def\etal{{\it et al.}}
%
\def\simle{\lower 2pt \hbox {$\buildrel < \over {\scriptstyle \sim }$}}
\def\simge{\lower 2pt \hbox {$\buildrel > \over {\scriptstyle \sim }$}}


\title{DARK MATTER AND STERILE NEUTRINOS}

\author{Peter L. Biermann$^{1,2,3}$ and   
Faustin Munyaneza\footnote{Humboldt Fellow}$^1$}
\address{
$^1$Max-Planck Institute for Radioastronomy, Bonn, Germany\\
$^2$Department of Physics and Astronomy,
University of Bonn,  Germany, \\
$^3$Department of Physics and Astronomy, University of Alabama,
Tuscaloosa, AL, USA
\email{plbiermann@mpifr-bonn.mpg.de, munyanez@mpifr-bonn.mpg.de}
}

\begin{abstract}

Dark matter has been recognized as an essential part of matter for over 70
years now, and many suggestions have been made, what it could be.  Most of 
these ideas have centered on Cold Dark Matter, particles that are predicted 
in extensions of standard particle physics, such as supersymmetry.  Here
we explore the concept that dark matter is sterile neutrinos, particles
that are commonly referred to as Warm Dark Matter.  Such particles 
have keV masses, and decay over a very long time, much longer than the 
Hubble time.  In their decay they produce X-ray photons which modify the
ionization balance in the very early universe, increasing the fraction
of molecular Hydrogen, and thus help early star formation.  Sterile
neutrinos may also help to understand the baryon-asymmetry, the pulsar
kicks, the early growth of black holes, the minimum mass of dwarf
spheroidal galaxies, as well as the shape and smoothness of dark matter
halos.  As soon as all these tests have been made quantitative in their various
parameters, we may focus on the creation mechanism of these particles,
and could predict the strength of the sharp X-ray emission line,
expected from any large dark matter assembly.  A measurement of this
X-ray emission line would be definitive proof for the existence of may
be called weakly interacting neutrinos, or WINs.

\end{abstract}

\keywords{Dark matter, sterile neutrinos, galaxies, black hole physics}
\bodymatter

\section{Introduction}

Since the pioneering works of Oort \cite{oort32} and 
Zwicky \cite{zwicky33,zwicky37},
 it has been known that there
is dark matter in the universe, matter that interacts gravitationally,
but not measurably in any other way.  Oort argued about the motion and
density of stars perpendicular to the Galactic plane, and in this case,
Oort's original hunch proved to be correct, the missing matter turned
out to be low luminosity stars.  Zwicky argued about the motions and
densities of galaxies in clusters of galaxies, and to this day clusters
of galaxies are prime arguments to determine dark matter, and its
properties \cite{vikhlinin06,pratt07}.

Based on the microwave back ground fluctuations\cite{spergel06}
today we know that the universe is flat geometrically, i.e. the sum of
the angles in a
cosmic triangle is always 180 degrees, provided we do not pass too close
to a
black hole.   This finding can be translated into stating that the sum
of the mass and energy contributions to the critical density of the
universe add up to unity, with about 0.04 in baryonic matter, about 0.20
in dark matter, and the rest in dark energy;  we note that there is no
consensus even where to find all the baryonic matter, but a good guess
is warm to hot gas, such as found in groups and clusters of galaxies,
and around early Hubble type galaxies\cite{viel05,cenostriker06}.
 
There are many speculations of what dark matter is; we have three
constraints:

1)  It interacts almost exclusively by gravitation, and not measurably in
any other way.  2)  It does not participate in the nuclear reactions in
the early universe. 3)  It must be able to clump, to help form galaxies,
and later clusters of galaxies, and the large scale structure\cite{bertone05}.
 Other options, such as particles in higher dimensions,
also exist\cite{biermannframpton06}, and other gravitational theories\cite{sandersmcgaugh02,
sanders05,dodelsonliguori06,carrol06,bekensteinsanders06}.
Obviously, various extensions in particle physics theory, such as
supersymmetry,  provide candidates like the lightest supersymmetric
particle.

Here we focus on the concept that it may be a ``sterile neutrino", a
right-handed neutrino, that interacts only weakly with other neutrinos,
and otherwise only gravitationally.  Such particles were proposed first
by Pontecorvo\cite{pontecorvo67} and later by Olive \& Turner \cite{oliveturner82}.
Sterile neutrinos were further proposed as 
  as dark matter
candidates\cite{dodelsonwidrow94}.
It was then shown  how oscillations of normal neutrinos to sterile neutrinos could
help explain the very large rectilinear velocities of some pulsars\cite{kusenkosegre97}.

Observationally the evidence comes from a variety of arguments:
i)  Dark matter in a halo like distribution is required to explain the
stability of spiral galaxy disks\cite{ostrikerpeebles73,ostrikerpeebles74,toomre77};
 ii) the flat rotation curves of galaxies\cite{rubin80};
  and iii)  the containment of hot gas in early Hubble type
galaxies\cite{biermann82}.  Dark matter is  also required to
explain: iv)  the structure of clusters of galaxies\cite{ensslim97,bohringer04};
 v) the  structure formation \cite{scannapiero06}
 and  the flat geometry of the universe\cite{spergel03,spergel06}.
We refer the reader to a recent 
review\cite{bertone05} and a book \cite{schneider06}
on dark matter.

Therefore after more than 70 years we still face the question: ``What is
dark matter?"

\section{The dark matter session at Berlin}

A much wider range of topics was explored in the session, than what we
can survey in extenso; we will confine ourselves to the limited topic at
hand subsequently.

\begin{itemize}
\item{} Shaposhnikov, Mikhail: Sterile neutrino dark matter. \\ His
lecture is also contained among the reports.  He reviewed the basic
particle physics ingredients and some of his latest work is summarized
in various papers \cite{asakashapo05, asaka05, asaka06a, asaka06b,
 bezrukovshapo06,shapotkachev06}.

 \item{} Mapelli, Michela: Impact of dark matter decays and
annihilations on reionization\\
We consider four different dark matter 
candidates (light dark matter, gravitinos, neutralinos and sterile
neutrinos), for each of 
them deriving the decaying/annihilation rate, the influence on
reionization, matter 
temperature and structure formation. We find that light dark matter
particles (1\;  - \; 10  MeV) and
 sterile neutrinos (2\;  - \;8  keV) can be sources of partial early
reionization (z$ \; <~ \; 100$). However,
  their integrated contribution to Thomson optical depth is small
($<~0.01$) with respect to the three year WMAP results. Finally, they
can significantly affect the behavior of matter temperature, delaying
the formation of first stars. On the contrary, effects of heavy dark
matter candidates (gravitinos and neutralinos) 
  on reionization and heating are minimal\cite{mapelliferrara05,mapelli06}.
   
\item{} Ripamonti, Emanuele: Dark matter decay and annihilation
influence upon structure formation\\
  DM decays and annihilations might heat and ionize the primordial IGM,
affecting its thermal and chemical evolution. We investigate whether
they can also change the "critical mass" which is needed for a
primordial halo to cool and form stars. This is done through a 1-D
hydrodynamical code, where we included the treatment of the chemical
evolution of the gas, and of the effects of different models of DM
decays/annihilations. The results are mixed: in some cases (e.g. sterile
neutrinos) the critical mass remains almost unchanged; in other cases
(e.g. light dark matter decays) it increases significantly. In fact, the
enhanced ionized fraction catalyzes H$_2$ formation and favours cooling,
but this is not sufficient to compensate for the effects (e.g. the
increase in the Jeans mass) of the extra heating \cite{ripamonti06}.

\item{} Stasielak, Jaroslaw: Evolution of the primordial clouds
in the warm dark matter model with keV sterile neutrinos\\
    We analyze the processes relevant for star formation in a model with
dark matter in the form of sterile neutrinos. Sterile neutrino decays
produce an X-ray background radiation that has a two-fold effect on the
collapsing clouds of hydrogen. First, the X-rays ionize gas and cause an
increase in the fraction of molecular hydrogen, which makes it easier
for the gas to cool and form stars. Second, the same X-rays deposit a
certain amount of heat, which could, in principle, thwart the cooling of
gas. We find that, in all the cases we have examined, the overall effect
of sterile dark matter is to facilitate the cooling of gas. Hence, we
conclude that dark matter in the form of sterile neutrinos can
facilitate the early collapse of gas clouds and the subsequent star
formation\cite{jarek06}.

\item{} Slosar, Anze:  Cosmological constraints on sterile neutrinos\\
       Sterile neutrinos come in two kinds in the cosmological context.
On one hand we have very weakly coupled sterile neutrinos with masses of
the order of a few keV that act as a dark matter candidate. On the other
hand people are also considering light and completely thermalized
neutrino species with masses of around 1 eV or less. I will discuss how
cosmology can constrain both of these sterile neutrino candidates, what
are the present limits and possible work-arounds\cite{mangano06,melchiorri06}.

\item{} Guzman, Murillo Francisco Siddhartha:  Scalar field dark matter:
beyond the spherical collapse\\
  We show the evolution of non-spherically symmetric balls of a
self-gravitating scalar field in the Newtonian regime or equivalently an
ideal self-gravitating condensed Bose gas. In order to do so, we use a
finite difference approximation of the Schr\"odinger-Poisson (SP)
system of equations with axial symmetry in cylindrical coordinates. Our
results indicate: 1) that spherically symmetric equilibrium
configurations are stable against non-spherical perturbations and 2)
that such equilibrium configurations of the SP system are late-time
attractors for non-spherically symmetric initial profiles of the scalar
field, which is a generalization of such behavior for spherically
symmetric initial profiles. Our system and the boundary conditions used,
work as a model of scalar field dark matter collapse after the
turnaround point. In such case, we have found that the scalar field
overdensities tolerate non-spherical contributions to the profile of the
initial fluctuation\cite{bernalguzman06a,bernalguzman06b}.

\item{} Khriplovich, Iosif:  Upper limits on density of dark matter in
Solar system\\
  The analysis of the observational data for the secular perihelion
precession of Mercury, Earth, and Mars, based on the EPM2004
ephemerides, results in new upper limits on density of dark matter in
the Solar system\cite{khriplovich06}.

\item{} Popa, Lucia and  Vasile, Ana:  Sterile Neutrino as Dark Matter 
candidate from CMB alone\\
Distortions of CMB temperature and polarization maps caused by gravitational
 lensing, observable with high angular resolution and sensitivity, can 
be used
 to constrain the sterile neutrino mass, $m_s$, from CMB data alone. We 
forecast
 $m_{s} >1.75$ keV from Planck and $m_{s} >4.97$ keV from Inflation Probe at 95\%
CL, by using the CMB weak lensing extraction\cite{popavasile07}.

\item{} Kronberg, Philip P.: Interconnections between black holes,
magnetic fields, cosmic rays in the Universe: A Review \\
I review the interconnections among these, focusing on consequences of
energy efficiency and the magnetic field generation by massive black
holes. The remarkable conversion efficiency from gravitational to
magnetic energy is discussed. Although it is ill-understood, recent
global tests already serve to constrain magnetic field creation scenarios.\\
Energy flows and magnetic fields arising from galactic black holes are
discussed, and also the close connections to cosmic ray acceleration
issues. New radio detections of diffuse extragalactic radio emission on
degree-scales will be described, and also the first attempts to probe
the level of magnetic fields in cosmological large scale structure
directly by Faraday rotation. These are compared with theoretical
predictions of diffuse magnetic field strengths\cite{kronberg06}.

\item{} Gergely, Laszlo: Is dark matter futile on the brane?\\
Rather than introducing various type of candidates for dark matter,
gravitational dynamics can be modified in order to explain the 
observations.
One route to do this is in the framework of the so-called brane-world
theories, \cite{RS2,MaartensLR} in which our observable universe is a 
brane
embedded into a higher dimensional space-time (the bulk). In these 
theories
the apparent gravitational dynamics on our observable $4$-dimensional
universe is given by the twice contracted Gauss equation, the Codazzi
equation and the effective Einstein equation.\cite{Decomp}

Gergely has discussed whether dark matter can be
replaced by various source terms appearing in the effective Einstein
equation. Such non-conventional source terms include a quadratic 
(ordinary)
matter source term, a geometric source term originating in the Weyl
curvature of the bulk, a source term arising from the possible asymmetric
embedding, and finally the pull-back to the brane of possible non-standard
model bulk fields.

The non-linear source term modifies only the very early 
cosmology\cite{BDEL},
due to the enormous value of the brane tension.

The Weyl curvature of the bulk in a spherically symmetric brane-world 
metric
generates a dark matter mass\cite{HarkoCheng}. With no cosmological
constant, the dark mass scales linearly with the radial distance, 
explaining
the flatness of the galactic rotation curves.

Properly chosen non-standard model bulk fields can replace dark matter in
explaining structure formation\cite{Pal}, the evolution of perturbations 
on
the brane becoming similar to that of the Cold Dark Matter (CDM) model.

With a radiating black hole only on one side of the brane, interesting
phenomena occur. The combined effect of asymmetry and a bulk radiation
qualitatively gives both dark matter and dark energy\cite{Irradiated,Irradiated2}.
 The radiation pressure accelerates the brane, as
would dark energy do, while the absorbed radiation increases the energy
density of the bulk, appearing as CDM. These two effects compete with each
other, and with properly chosen initial data a critical-like behavior can 
be
found.

\item{} Watson, Casey:  Direct X-ray Constraints on Sterile Neutrino
Warm Dark Matter\\
  In this talk, I will discuss how we use the diffuse X-ray spectrum 
   (total minus resolved point source emission) of the Andromeda galaxy
to constrain the rate of sterile neutrino radiative decay and the
sterile neutrino mass, $m_s$. Our findings demand that  $m_{s} < 3.5$
keV (95\% C.L.) which is a significant improvement over the previous
(95\% C.L.) limits inferred from the X-ray emission of nearby clusters,
$m_s < 8.2$ keV (Virgo A) and $m_s < 6.3$ keV (Virgo A + Coma) \cite{watson06}.
    
\item{} Riemer-S\mbox{\o}rensen, Signe:  Probing the nature of dark
matter with Cosmic X-rays\\
Gravitational lensing observations of galaxy clusters have identified
dark matter ``blobs'' with remarkably low baryonic content. We use such
a system to probe the particle nature of dark matter with X-ray
observations. We also study high resolution X-ray grating spectra of a
cluster of galaxies. From these grating spectra we improve the
conservative constraints on a particular dark matter candidate, the
sterile neutrino, by more than one order of magnitude. Based on these
conservative constraints obtained from cosmic X-ray observations alone,
the low mass ($m_s < 10$ keV) and low 
mixing angle $(\sin^2 (2\theta) <
10^{-6}) $ sterile neutrino is still a viable dark matter
 candidate\cite{riemer06a,rimer06b}.
 In Fig.~1, the lifetime of the sterile neutrino is plotted as a function
 of the photon energy $E$ in keV.

\item{} Munyaneza, Faustin: Limits on the dark matter particle mass 
from black hole growth in galaxies\\
  I review the properties of degenerate fermion balls and investigate
the dark matter distribution at galactic centers using NFW, Moore and
isothermal density profiles. I show that dark matter becomes degenerate
for particles masses of a few keV and for distances less than a few
parsec from the center of our galaxy. To explain the galactic center
black hole of mass of $\sim 3.5 \times 10^{6}M_{\odot}$ and a
supermassive black hole of $\sim 3 \times 10^{9}M_{\odot}$ at a redshift
of 6.41 in SDSS quasars\cite{willot03},
 we
require a fermion ball mass between $3 \times 10^{3} M_{\odot}$ and $3.5
\times 10^{6}M_{\odot}$. This leads to strong limits on the mass of the
dark matter particle between $0.64 \ {\rm keV}$ and $5.82 \ {\rm keV}$
for NFW profile, and between $0.97 \ {\rm keV}$ and $13.81 \ {\rm keV}$
for Moore profile and finally the dark matter mass is found to be
constrained between $2.38 \ {\rm keV}$ and $ 81.65 \ {\rm keV}$ for the
isothermal gas sphere case. I then argue that the constrained particle
could be the long sought dark matter of the Universe that is interpreted
as a sterile 
neutrino\cite{munyanezabiermann05,munyanezabiermann06,munyanezabiermanngc06}.
The limits on the sterile neutrino mass  are shown in
Fig.~2.  

\item{} Ruchayskiy, Oleg: Search for the light dark matter with an X-ray
spectrometer\\
Sterile neutrinos with the mass in the keV range are interesting warm
dark matter (WDM) candidates. The restrictions on 
 their parameters (mass
and mixing angle) obtained by current X-ray missions (XMM-Newton or
Chandra) can only be improved by less than an order of magnitude in the
near future. Therefore the new strategy of search is needed. We compare
the sensitivities of existing and planned X-ray missions for the
detection of WDM particles with the mass  1 - 20 keV. We show that
existing technology allows an improvement in sensitivity by a factor of
100. Namely, two different designs can achieve 
such an improvement: [A]
a spectrometer with the high spectral resolving power of 0.1 \%,
wide (steradian) field of view, with small effective area of about
$cm^2$ (which can be achieved without focusing optics) or [B] the same
type of spectrometer with a smaller (degree) field of view but with a
much larger effective area of $10^3 cm^2$ (achieved with the help of
focusing optics). To illustrate the use of the "type A" design we
present the bounds on parameters of the sterile neutrino obtained from
analysis of the data taken by an X-ray microcalorimeter. In spite of the
very short exposure time (100 sec) the derived bound is comparable to
the one found from long XMM-Newton observation\cite{boyarsky06d}.

\item{} Gilmore, Gerard: Dwarf spheroidal galaxies\\
The Milky Way satellite dwarf spheroidal (dSph) galaxies are the
smallest dark matter dominated systems in the universe. We have underway
dynamical studies of the dSph to quantify the shortest scale lengths on
which Dark Matter is distributed, the range of Dark Matter central
densities, and the density profile(s) of DM on small scales. Current
results suggest some surprises: the central DM density profile is
typically cored, not cusped, with scale sizes never less than a few
hundred pc; the central densities are typically 10-20 GeV/cc; no galaxy
is found with a dark mass halo less massive than $~ 5 \; 10^{7}$ ${\rm
M_{\odot}}$. We are discovering many more dSphs, which we are analysing
to test the generality of these results\cite{gilmore06}.
In Fig.~3, the inferred dark mass is shown for different dwarf spheroidal galaxies.
\end{itemize}

\begin{figure}
\includegraphics[width=12cm,height=10cm]{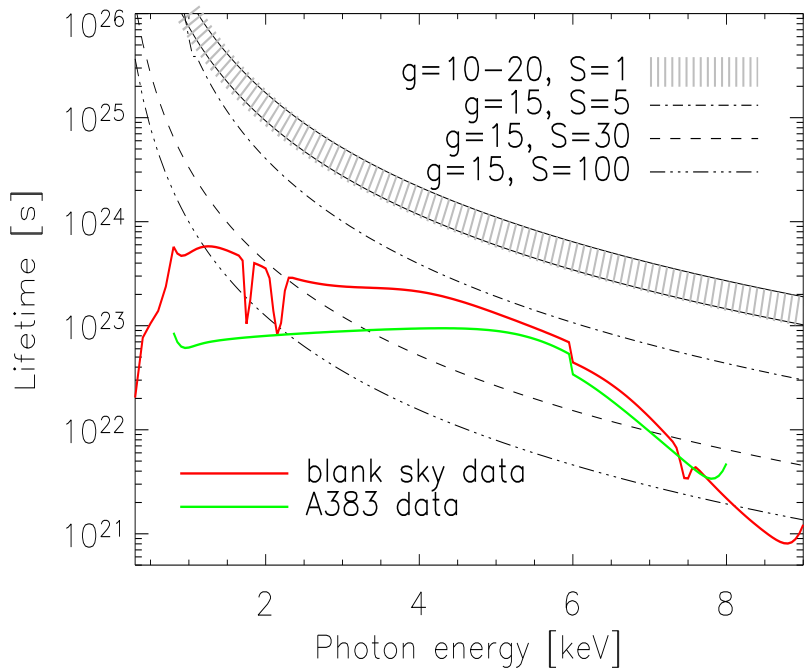}
\caption{The lifetime constrained from the flux of the Chandra blank sky data (red)
and A383 (green). The $\nu$MSM prediction for $S=1$ and $g_{*}=10-20$ (hatched) and
several variations of $S$ and $g_{*}$ (black) have been overplotted. This plot has been
adapted from ref.~\cite{riemer06a} under the permission of the authors. Full explanation of the
parameters $S$ and $g_{*}$ are found in the same reference.}
\end{figure}

\begin{figure}
\includegraphics[width=12cm,height=10cm]{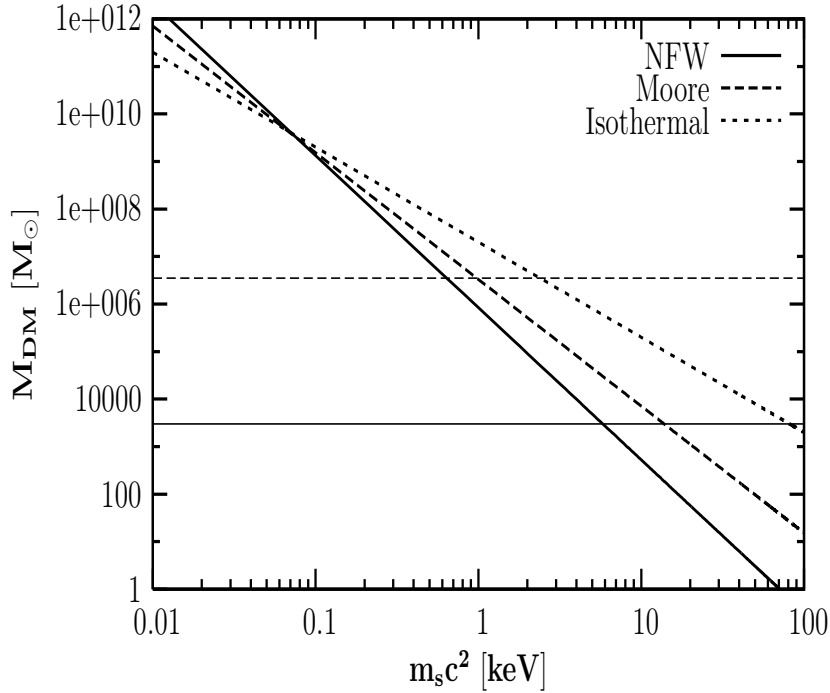}
\caption{ The total mass $M_{DM}$ of the  fermion ball as a function of the
 fermion mass $m_{s}$. 
The total mass $M_{DM}$ scales with the the DM particle mass $m_{s}$ as $m_{s}^{-16/5}$,
$m_{s}^{-8/3}$ and as $m_{s}^{-2}$
for NFW, Moore and isothermal power law, respectively.
 Two horizontal lines 
at $M_{DM}=3.5 \times 10^{6} M_{\odot}$ and $M_{DM}=3 \times 10^{3}M_{\odot}$ have been drawn 
 to get the lower and upper limits on the mass of the DM particle. From this plot,
 a sterile neutrino mass could be in the range from 1 keV to about 80 keV.
 The lower limit could be improved to about 7 keV  if the Fermi Dirac distribution
 is used in solving the Poisson's equation for the gravitational
 pontential of sterile neutrinos\cite{munyanezabiermann05,munyanezabiermanngc06}.
  This graph has been  taken from ref. \cite{munyanezabiermann06}.}
\end{figure}

\begin{figure}
\includegraphics[width=12cm,height=10cm]{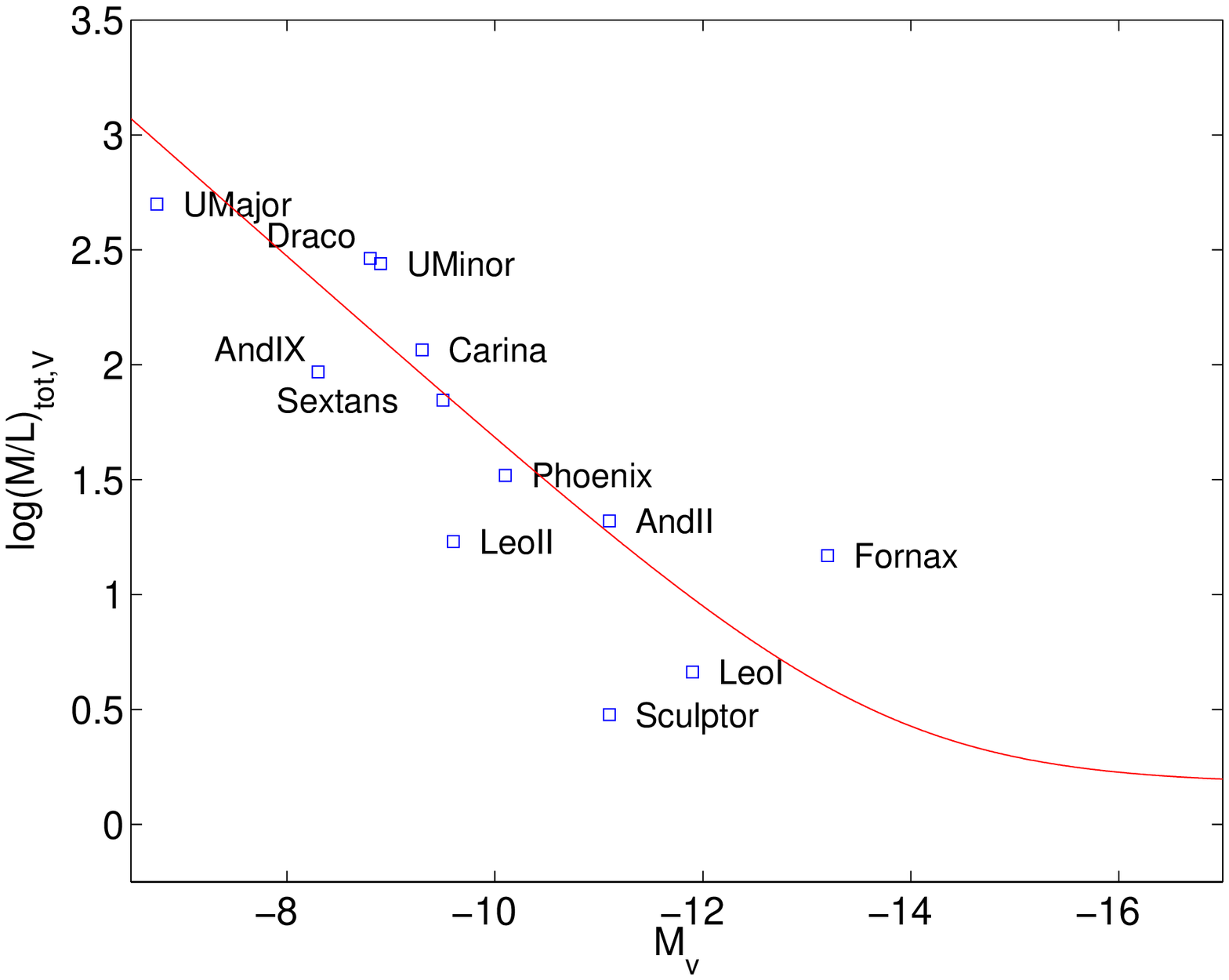}
\vspace{1.2cm}
\caption{Mass to light ration vs galaxy absolute
V magnitude for some Local Group dSphs galaxies.
The solid curve shows the relation expected
if all the dSphs galaxies contain about
$4 \times 10^{7}M_{\odot}$ solar masses of dark 
matter interior to their stellar distributions.
This plot has been adapted from ref.~\cite{gilmore06} under
the permission of the author.}
\end{figure}

\section{Proposal}

The existing proposals to explain dark matter mostly focus on very massive
particles\cite{bertone05},
 such as the
lightest
supersymmetric particle; all the experimental searches are sensitive for
masses above GeV, usually far above such an energy.  In the normal
approach to structure formation, this implies a spectrum of dark matter
clumps extending far down to globular cluster masses and below.  It has
been a difficulty for some time that there is no evidence for a large
number of such entities near our Galaxy.  The halo is clumpy in stars,
but not so extremely clumpy.  If, however, the mass of the dark matter
particle were in the keV range, then the lowest mass clumps would be
large enough to explain this lack.  However, in this case the first star
formation would be so extremely delayed\cite{yoshida03}
 that there
would be no explanation of the early reionization of the universe,
between redshifts 11 and 6, as we now know for sure\cite{spergel03,spergel06}.
Therefore, the conundrum remained.

Here we explore the concept that the dark matter is indeed of a mass in the
keV range, but can decay, and so produce in its decay a photon, which
ionizes.  It so increases the abundance of molecular Hydrogen, and so
allows star formation to proceed early \cite{jarek06,biermannkusenko06}.
  The specific model we explore is
of ``sterile neutrinos", right handed neutrinos, which interact only
with normal, left-handed neutrinos, and with gravity.  Such particles
are commonly referred to as ``Warm Dark Matter", as opposed to ``Cold
Dark Matter", those very massive particles. For most aspects of
cosmology warm dark matter and cold dark matter predict the same; only
at the small scales are they significantly different, and of course in
their decay.

The mass range we explore is approximately 2 - 25 keV.  These sterile
neutrinos decay, with a very long lifetime, and in a first channel give
three normal neutrinos, and in the second channel, a two-body decay,
give a photon and a normal neutrino. The energy of this photon is almost
exactly half the mass of the initial sterile neutrino.

What is important is to understand that such particles are not produced
from any process in thermal equilibrium, and so their initial phase space
distribution is far from thermal; current models for their
distribution suggest that their momenta are sub-thermal. The measure of
how much they are subthermal modifies the precise relationship between
the dark matter particle mass and the minimum clump mass, which should
be visible in the smallest pristine galaxies.

This also entails, that as Fermions they require a Fermi-Dirac
distribution, as
being far from equilibrium, this distribution implies a
 chemical potential\cite{munyanezabiermann05}.

Recent work by many others \cite{abazajian01, abazajian05, abazajiankoushiapas06,
abazajian06a,abazajian06b, boyarsky06a,boyarsky06b,boyarsky06c, dolgovhansen02, 
asakashapo05, asaka05,
asaka06a,asaka06b,bezrukovshapo06,shapotkachev06}
  has shown that these sterile neutrinos
can be produced in the right amount to explain dark matter, could explain
the baryon asymmetry \cite{akhmedov98}, explain the lack of power on
small scales (as noted above), and could explain the dark matter
distribution in galaxies \cite{belokurov06,fellhauer06,gilmore06}.

\subsection{Our recent work}

Pulsars are observed to reach linear space velocities of up to over 1000
km/s, and there are not many options how to explain this; one
possibility is to do this through magnetic fields which become important
in the explosion\cite{bisnovatyi70, bisnovatyi93,bisnovatyi95,ardelian05,moiseenko05};
  curiously, normal neutrinos played a role already in these
early ideas.  Another possibility is to do this through a conversion of
active neutrinos which scatter with a mean free path inside the
exploding massive star of about ten cm  \cite{shapiroteukolsky83},
 into
sterile neutrinos, which no longer scatter.  If this conversion produces
a spatial and directional correlation between the sterile neutrinos and
the structure of the highly magnetic and rotating core of the exploding
star, then a small part of the momentum of the neutrinos can give an
asymmetric momentum to the budding neutron star, ejecting it at a high
velocity\cite{kusenko04}.  This then could explain such features as the
guitar nebula\cite{chatterjee04}, the bow shock around a high
velocity pulsar.  This latter model in one approximation requires a
sterile neutrino in the mass range 2 to 20 keV. It is remarkable that
this neutrino model requires magnetic fields in the upper range of the
strengths predicted by the magneto-rotational mechanism to explode
massive stars as supernovae.

 It was shown  from
SDSS data \cite{fan01,willot03}, that some quasars have supermassive black holes already at
redshift 6.41, i.e. at about  800 million years after the big bang.  We now know,
that this is the period when galaxies grow the fastest, from 500 to 900
million years after the big bang\cite{bouwens06,iye06}.
 Baryonic accretion has trouble feeding a normal black hole
to this high mass, $ 3 \; 10^{9}$ solar masses so early after the big
bang, if the growth were to start with stellar mass black holes\cite{wangbiermann98}.
  So either the first black holes are around
$10^{4}$ to $10^{6}$ solar masses, and there is not much evidence for
this, or the early black holes grow from dark 
matter\cite{munyanezabiermann05,munyanezabiermann06,munyanezabiermanngc06},
 until they reach the critical minimum mass to be
able to grow very fast and further from baryonic matter, which implies a
mass range of about $10^{4}$ to $10^{6}$ solar masses.  This model in
the isothermal approximation for galaxy structure implies a sterile
neutrino in the mass range between 12 and 450 keV.

When Biermann and Kusenko met at an Aspen meeting September 2005,
it became apparent, that these two speculative but very different
approaches overlap, and so it seemed worth to pursue them further.

As noted above, structure formation arguments lead to an overprediction
in power at small scales in the dark matter distribution in the case of
cold dark matter, and any attempt to solve this with warm dark matter
delayed star formation unacceptably\cite{yoshida03}. We convinced
ourselves that this was the key problem in reconciling warm dark matter
(keV particles) with the requirements of large scale structure and
reionization.
We then showed that the decay of the sterile neutrino could
increase the ionization, sufficiently to enhance the formation of
molecular hydrogen, which in turn can provide catastrophic cooling early
enough to allow star formation as early as required \cite{biermannkusenko06,jarek06}.
In our first simple calculation this happens at redshift 80.  More refined
calculations confirm, that the decay of sterile neutrinos helps increase
the fraction of molecular Hydrogen, and so help star formation, as long as
this is at redshifts larger 
than about 20 \cite{mapelliferrara05, mapelli06,ripamonti06}.

\section{The tests}

\subsection{Primordial magnetic fields }

In the decay a photon is produced, and this photon ionizes Hydrogen: at
the first ionization an energetic electron is produced, which then
ionizes much
further, enhancing the rate of ionization by a factor of about 100.  In
the case, however, that there are primordial magnetic fields, this
energetic electron could be caught up in wave-particle interaction, and
gain energy rather than lose energy.  As the cross section for
ionization decreases with energy, the entire additional ionization by a
factor of order 100 would be lost in this case, and so there basically
would be no measurable effect from the dark matter decay.  This gives a
limit for the strength of the primordial magnetic field, given various
models for the irregularity spectrum of the field:  In all reasonable
models this limit is of order a few to a few tens of picoGauss,
recalibrated to today.

Recent simulations  matched to the magnetic field
data of clusters and superclusters, give even more stringent limits, of
picoGauss or less \cite{dolag05}.

It follows that primordial magnetic fields can not disturb the early
ionization from the energetic photons, as a result of dark matter decay.
 It then also follows that the contribution of early magnetic fields
from magnetic monopoles, or any other primordial mechanism, is
correspondingly weak\cite{wick03}.

Stars at all masses are clearly able to produce magnetic
 fields\cite{biermannl50, biermannl51, silklanger06},
 but the evolution and consequent dispersal are fastest for the massive
stars, almost certainly the first stars.  As the magnetic fields may
help to drive the wind of these massive stars\cite{seemanbiermann97},
 then the wind is just weakly super-Alfv\'enic, with Alfv\'enic
Machnumbers of order a few.  This implies that the massive stars and
their winds already before the final supernova explosion may provide a
magnetic field which is at order 10 percent equipartition of the
environment; this magnetic field is highly structured.  However, even
these highly structured magnetic fields will also allow the first cosmic
rays to be produced, and distributed, again with about 10 percent of
equipartition of the environment.  The question on how magnetic fields arising from
galaxies can  get distributed across the cosmos has been investigated in 
by various authors \cite{ryu98,ensslim98, gopal04,kronberg06}.
However, the large scale structure and coherence of the cosmic magnetic
fields clearly remain an unsolved problem \cite{kulsrud99,biermanngalea03,biermannkronberg04}.
 
Therefore the first massive stars are critical for the early evolution of
the universe:  In addition to reionization, magnetic fields and cosmic
rays, these stars  provide the first heavy elements.  These heavy elements allow
in turn
dust formation, which can be quite rapid (as seen, e.g., in SN 1987A,
already
just years after the explosion \cite{biermann90}).  This then
enhances the cooling in the dusty regions, allowing the next generation
of stars to form much faster.

In combination everywhere one first massive star is formed, we can
envisage a runaway in further star formation in its environment.

\subsection{Galaxies}

Galaxies merge, and simulations demonstrate that the inner dark matter
distribution attains a power law in density, and a corresponding power
law tail in the momentum distribution \cite{nfw97,moore99,klypin02}:
 Here the central density
distribution as a result of the merger is a divergent power law, as a
result of energy flowing outwards and mass flowing inwards, rather akin
to accretion disks \cite{lust52,lustschluter55,prendergast68,shakura72,
shakura73,lyndenbel74},
  where angular momentum flows outwards and
mass also flows inwards; in fact also in galaxy mergers angular momentum
needs to be redistributed outwards as such mergers are almost never
central\cite{tomre72}.  This then leads to a local escape
velocity converging with $r$ to zero, and so for
fermions the Pauli limit is reached, giving rise to a cap in density,
and hence give birth to a dark matter star, also called a
 fermion ball\cite{munyanezabiermann05,munyanezabiermann06}.
this dark matter star can grow further by dark matter accretion.  
The physics of fermion balls  as a model for the 
 dark matter distribution 
at galactic centers
has been  studied in a series of papers\cite{viollier94,tv98,mtv98,mtv99,bmv99,mv02,bmtv02}.
For
realistic models an integral over a temperature distribution is
required, and a boundary condition has to be used to represent the
surface of the dark matter star both in real space as in momentum space.
 This then allows the mass of this dark matter star to increase further.
 In quantum statistics,
such models resemble  white dwarf stars or
neutron stars as  the Pauli's pressure upholds the star.
 As seen in Fig.~2, for fermions in the
keV range,  the mass of the dark matter star has a mass range of a few
thousand to a few million solar masses.

The first stellar black hole can then enter this configuration and eat
the dark matter star from inside, taking particles from the low angular
momentum phase space.  With phase space continuously  being refilled through
the turmoil of the galaxy merger in its abating stages, or in the next
merger, the eating of the dark matter star from inside ends only when
all the dark matter star has been eaten up.

Given a good description of the dark matter star boundary conditions in
real and in momentum phase space\cite{munyanezabiermann05,munyanezabiermann06},
 and an observation of the stellar velocity dispersion close to
the final black hole, but outside its immediate radial range of
influence, we should be able to determine a limit to the dark matter
particle mass.  If the entire black hole in the Galactic Center has
grown from dark matter alone, then we obtain a real number of about
6 to 10 keV \cite{munyanezabiermann05,munyanezabiermanngc06}.

This concept suggests that it might be worthwhile to consider the
smallest of all black holes in galactic centers.  In a plot of black
hole mass $M_{BH}$ versus central stellar velocity dispersion $\sigma$
 there is a clump above the
relation  $M_{BH} \; \sim \; \sigma^{4}$ at low black hole masses\cite{barth05,greene06},
suggesting that perhaps we reach a limiting relationship with a flatter
slope for all those black holes which grow only from dark matter. For an
simple isothermal approach this flatter slope is 3/2 as obtained in ref. \cite{munyanezabiermann06}.

\subsection{Dwarf spheroidal galaxies}

 All detected dwarf spheroidal galaxies
fit a simple limiting relationship of a common dark matter mass of $4 \,
10^7$ solar masses\cite{gilmore06}, suggesting that this is perhaps the
 smallest dark
matter clump mass in the initial cosmological dark matter clump
spectrum.  This clump mass is of course a lower limit to the true
original mass of the pristine dwarf spheroidal galaxy.
Fig.~3 shows the distribution of dark matter mass for various
dwarf spheroid galaxies.
  Given a physical
concept for the production of the dark matter particles in the early
universe, we would have their initial momentum, probably subthermal, and
so the connection between the dark matter particle mass and minimum
clump mass is modified.  This is a very strong support for the Warm Dark
Matter concept and the associated problems have been pointed out in 
a number of papers \cite{metz06,sohn06,strigari06}.

One intriguing aspect of dwarf spheroidal galaxies is that almost all of
them show the effect of tidal distortion in their outer regions, and at
least one of them has been distended to two, perhaps even three
circumferential rings around our Galaxy\cite{belokurov06,fellhauer06}.
  To extend so far around our Galaxy must have
taken many orbits, and so some fraction of the age of our Galaxy.  The
simple observation that these streamers still exist separately, and can
be distinguished in the sky, after many rotations around our Galaxy,
implies that the dark matter halo is extremely smooth, and also nearly
spherically symmetric. Given hat the stellar halo is quite clumpy this
implies once more that the dark matter is much more massive than the
baryonic matter in our halo.
Metallicities and detailed spectroscopy give a throve of further
information\cite{koch06,koch07}.
Small spiral galaxies may or may not be different\cite{milgromsanders06}.

\subsection{Lyman alpha forest}

In the early structure formation the large number of linear
perturbations in density do not lead to galaxies, but just to small
enhancements of Hydrogen density, visible in absorption against a
background quasar.  This so-called Lyman alpha forest tests the section
of the perturbation scales which is linear and so much easier to
understand, and it should in principle allow a test for the smallest
clumps\cite{viel06,seljak06}.  Unfortunately,
systematics make this test still difficult, and with the expected
sub-thermal phase space distribution of the dark matter particles, we may
lack yet the sensitivity to determine the mass of the smallest clumps.

\begin{figure}
\includegraphics[width=12cm,height=10cm]{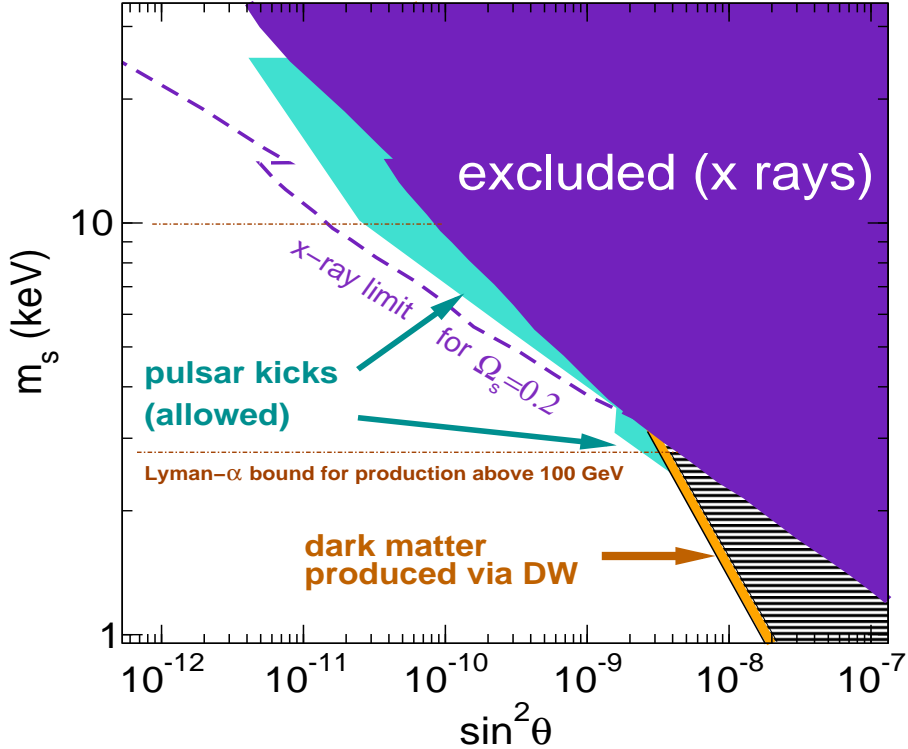}
\caption{ The X-ray limits reported in ref.~ \cite{riemer06a,boyarsky06c,abazajian01}
 (dashed line) apply if the sterile neutrinos
account for all the dark matter ($\Omega_{s}=0.2)$. The value of $\Omega_{s}$ depends
on the production mechanism, but it cannot be lower than the amount
 produced via the Dodelson Widrow
 mechanism
\cite{dodelsonwidrow94} (except for the lower-reheat 
scenarios \cite{gelmini04}). The model 
independent exclusion plot (purple region)
is obtained by assuming this minimal value. A sterile neutrino with mass 3 keV and
$sin^{2} \theta \approx 3 \times 10^{-9}$, produced at some temperature
 above 100 GeV, can explain
both pulsar kicks and dark matter.
This graph was taken from ref.\cite{kusenko06}
under the permission of the author.}
\end{figure}

\subsection{The X-ray test}

When the sterile neutrinos decay, they give off a photon with almost
exactly half their mass in energy.  Our nearby dwarf spheroidal
galaxies, our own inner Galaxy, nearby massive galaxies like M31, the
next clusters of galaxies like the Virgo cluster, and other clusters
further away, all should show a sharp X-ray emission line
\cite{abazajian01,riemer06a,rimer06b,watson06,boyarsky06b,boyarsky06d}.
Fig.1 shows the lifetime of the sterile neutrino for different photon energies. 
The universal X-ray background should show such a sharp emission line as
a wedge, integrating to high redshift.
With major effort this line or wedge should be detectable with the
current Japanese, American or European X-ray satellites: Large field
high spectral resolution spectroscopy is required.
Sterile neutrino mass limits are shown in Fig.~4.

\section{Outlook}

The potential of these right handed neutrinos is impressive, but in all
cases we have argued, there is a way out, in each case there is an
alternative way to interpret the data set.  E.g., for the pulsar kick
with the help of neutrinos strong magnetic fields are required, but the
MHD simulations suggest that perhaps magnetic fields can do it by
themselves, even without the weakly interacting neutrinos
 \cite{bisnovatyi70,bisnovatyi93,bisnovatyi95,ardelian05,moiseenko05}.
  The dwarf spheroidal galaxies can in some models be explained
without any dark matter at all\cite{metz06,sohn06}. 
The early growth of black holes can also be fuelled by other black
holes, as long as there are enough in number and their orbital angular
momentum can be removed.  So many alternatives may replace the sterile
neutrino concept.

However, the right handed, sterile neutrinos weakly interacting with the
normal left handed neutrinos provide a unifying simple hypothesis, which
offers a unique explanation of a large number of phenomena, so by
Occam's razor, it seems quite convincing at present\cite{kusenko06}.  So,
given what sterile neutrinos may effect, we may have to call them Weakly
Interacting Neutrinos, or soon WINs.

\section{Acknowledgements}
The authors wish to thank first and foremost Alex Kusenko who
 played a key role in working out  the science reported here.
PLB would  also like to acknowledge fruitful discussions with Kevork
Abazajian, Gennadi Bisnovatyi-Kogan, Klaus Dolag, Torsten En{\ss}lin,
Gerry Gilmore, Hyesung Kang, Phil Kronberg, Pavel Kroupa, Sergei
Moiseenko, Biman Nath, Dongsu Ryu, Mikhail Shaposhnikov, Jaroslaw Stasielak, Simon Vidrih,
Tomaz Zwitter and many others.

Support for PLB is coming from the AUGER membership and theory grant 05
CU 5PD 1/2 via DESY/BMBF. 
 FM's research is supported by the Alexander von Humboldt Foundation.

\end{document}